\documentclass [smallextended]{svjour3}
\usepackage{amssymb,amsmath,amsfonts,mathtools,bigints}
\allowdisplaybreaks 
\usepackage{cancel} 
\usepackage{mathrsfs}
\usepackage{graphicx} 
\usepackage{epstopdf}
\usepackage{pdflscape}
\usepackage[inline]{enumitem}
\usepackage{import}
\usepackage{aas_macros}
\usepackage{hyperref}
\usepackage[T1]{fontenc}
\usepackage{float}
\usepackage{fancyvrb} 
\usepackage{subfig}
\usepackage{array}
\usepackage{multirow} 
\usepackage{tabu} 
\usepackage[section]{placeins} 

\DeclareSymbolFontAlphabet{\mathrsfs}{rsfs}      
\DeclareMathOperator{\e}{\mathrm{e}}             

\DeclareMathOperator{\acoth}{arcoth}

\renewcommand{\d}{\ensuremath{\,\mathrm{d}}}     
\newcommand{\rmd}{\ensuremath{\mathrm{d}}}
\newcommand{\rme}{\ensuremath{\mathrm{e}}}

\newcommand{\f}[2]{\ensuremath{\frac{#1}{#2}} }  

\newcommand{\un}[1]{\ensuremath{\,\mathrm{#1}} }

\newcommand{\deriv}[2]{\ensuremath{\frac{\mathrm{d} #1}{\mathrm{d} #2}}}
\newcommand{\sderiv}[2]{\ensuremath{\frac{\mathrm{d}^2 #1 }{\mathrm{d} #2^2 }}}

\newcommand{\ppen}{\ensuremath{p_{\perp}}}

\let\oldsqrt\sqrt
\def\sqrt{\mathpalette\DHLhksqrt}
\def\DHLhksqrt#1#2{%
\setbox0=\hbox{$#1\oldsqrt{#2\,}$}\dimen0=\ht0
\advance\dimen0-0.2\ht0
\setbox2=\hbox{\vrule height\ht0 depth -\dimen0}%
{\box0\lower0.4pt\box2}}

{\\ \hline \\ \end{tabular} \end{center}}

\makeatletter
\renewcommand*\env@matrix[1][c]{\hskip -\arraycolsep
  \let\@ifnextchar\new@ifnextchar
  \array{*\c@MaxMatrixCols #1}}
\makeatother

\def\apjl{ApJL }

\def\apjs{ApJS }

\def\aap{A\&A }

\begin{document}
\title{Self-gravitating charged fluid spheres with anisotropic pressures} 
\author{Ambrish M. Raghoonundun \and
  David W. Hobill} \institute{Ambrish M. Raghoonundun \at University
  of Calgary, 2500 University Dr. NW, Calgary, Alberta, Canada, T2N
  1N4, \\\email{amraghoo@ucalgary.ca} \and David W. Hobill \at
  University of Calgary, 2500 University Dr. NW, Calgary, Alberta,
  Canada, T2N 1N4, \\\email{hobilll@ucalgary.ca} } \date{4 March
  2016}

\maketitle
\begin{abstract}
  The coupled system of the spherically symmetric Einstein--Maxwell
  differential equations is solved under two different source
  conditions: non-zero electric charge and pressure anisotropy.
  Expressions for the metric functions, and pressures which extend the
  Tolman~VII exact solution are deduced from the new solutions.  By
  applying boundary conditions to these solutions, all integration
  constants are computed in terms of parameters that are physically
  meaningful.

\end{abstract}

\section{Introduction}\label{intro}
The Tolman~VII solution~\cite{Tol39} is an exact analytic solution to
the static and spherically symmetric Einstein's field equations (EFE),
with a perfect fluid matter source.  This solution obeys all the
criteria for physical acceptability~\cite{DelLak98,FinSke98}, and has
been shown to be a viable solution that can be used for modelling
compact objects such as neutron and quark stars~\cite{RagHob15}.

In this article, that solution is generalized to generate new exact
solutions with charge and/or anisotropic pressures.  Since the
Tolman~VII solution provides a physically valid model of various
physical objects, the hope behind these generalizations is that they
will lead to new models that share these characteristics.

First a particular uncharged generalization of the Tolman~VII solution
is carried out by introducing an anisotropic pressure.  The case for
anisotropic pressures was made some time ago in
Refs.~\cite{HerSan97,BowLia74} and since then a number of studies into
their properties have been published.  Of note is the work of
Letelier~\cite{Let80} who elucidates the possible physical
interpretations of anisotropic pressures, and concludes that a mixture
of a number of perfect fluids can be transformed, through a coordinate
change, into \emph{one} anisotropic fluid.  This aspect simplifies the
physical interpretation of anisotropic pressure models, and leads one
to believe that anisotropic solutions can be candidates for simple
physical models of gravitating fluid spheres.

Under the assumption of pressure anisotropy, the components of the
pressure, which are assumed to be the same in all directions in a
perfect fluid, must now be generalized to two different functions,
which for intuitive reasons are usually called \(p_{r}\)
for the radial pressure component, and \(\ppen\)
for the angular pressure component.

The further generalization carried out in this article is the
introduction of electric charge in the models.  While the existence of
electrically charged, large scale physical objects are
unlikely~\cite{Cho08}, on time scales much shorter than the whole
lifetime of the compact object, charge could account for non-permanent
properties such as magnetism (for example to model magnetars) in these
objects.

Furthermore, as has been noted by numerous
authors~\cite{Iva02,KylMar67,VarRahRay10}, in the static limit, the
addition of charge does not change the difficulty of solving the EFE,
since a ``Maxwell differential equation'' for the electric charge is
added to the system of equations to be solved.  This can immediately
be integrated and incorporated into a global charge that is seen from
the outside only through the Reissner-Nordstr\"om external metric.
The EFE's do not change drastically either, and a similar solution
procedure to the one usually employed for the Tolman solutions can be
used to great effect.

As a guide to both old and new solutions, we provide an overview in
the form of a chart in Figure~\ref{ns.fig:SolFlow}.  This chart is
organized so that when moving from left to right, one moves from
solutions with isotropic pressures to those with anisotropic pressures
in the middle to solutions with both charge and anisotropic pressures
on the extreme right.  This Figure uses quantities that are explained
throughout the article.  Boxes that form an end point in the decision
flow are solutions that are either previously known (references given)
or are new and discussed in the section indicated.

This article is organized as follows. Section~\ref{ns.sec:ODE}
introduces the system of equations to be solved, and goes through the
process of solving the system, while providing physical motivations
for the various assumptions required for the solution.
Section~\ref{ns.sec:Ani} concentrates on uncharged anisotropic
pressures, and Section~\ref{ns.sec:Cha} constructs new solutions which
have electric charge. The last section reviews the methods and
discusses further research to be carried out on the physical relevance
of the solutions.

\section{The ODE system,  method of solution, and physical considerations}
\label{ns.sec:ODE}
\subsection{The ODE system}
In what follows, the metric is assumed to be spherically symmetric,
static, and expressed in the usual Schwarzschild spherical coordinates
\((t, r, \theta, \varphi)\) as
\begin{equation}\label{eq:metric}
\rmd s^{2} = \rme ^{\nu(r)} \rmd t^{2} - \rme ^{\lambda(r)} \rmd
r^{2} - r^{2} \rmd \theta^{2} - r^{2} \sin^{2} \theta \rmd
\varphi^{2}.  
\end{equation}
Throughout this derivation, an equivalent re-parametrization of the
metric coefficients is also used, so that
\(Z(r) = \rme^{-\lambda(r)}\) and \(Y(r) = \rme^{\nu(r)/2}.\) This
re-parametrization will allow the differential equations to be
simplified and depending on the type of matter considered in the
following sections, the EFE can be reduced to linear ODEs for \(Z(r)\)
and \(Y(r).\)

The matter content is specified by the form of the energy momentum
tensor.  The spherically symmetric fluid energy--momentum tensor with
anisotropic pressure is given by
\begin{equation}
\label{ns.eq:EMA}
  T^{i}{}_{j} =
  \begin{pmatrix}[c]
    \rho   & 0  & 0  & 0  \\
    0     & -p_{r}  & 0  & 0  \\
    0     & 0  & -\ppen & 0  \\
    0     & 0  & 0  & -\ppen 
  \end{pmatrix},
\end{equation}
with \(\rho, p_{r}\)
and \(\ppen\)
the matter density, radial pressure, and angular pressure,
respectively.

The addition of electric charge to this system is made through the
Faraday tensor, which in the static case is given~\cite{MTW,Cho08} by
\begin{equation}
  F_{ab} =
  \begin{pmatrix}[c]
    0     & -\f{qY}{r^{2}\sqrt{Z}}  & 0  & 0  \\
    \f{qY}{r^{2}\sqrt{Z}}     & 0  & 0  & 0  \\
    0     & 0  & 0  & 0  \\
    0     & 0  & 0  & 0
  \end{pmatrix}.
\label{ns.eq:Faraday}
\end{equation}
Here \(q(r)\) is the charge enclosed, with \(Y\) and \(Z\) the metric
functions previously defined.  The electromagnetic energy momentum
tensor is of the form
\[T^{\text{EM}}_{ab} = g_{ac}F^{cd}F_{db} - \f{1}{4} g_{ab}
  F^{cd}F_{cd}, \] and together with~\eqref{ns.eq:Faraday}
and~\eqref{ns.eq:EMA}, the total energy momentum tensor for the fluid
and the electric field source is given by~\cite{Iva02}:
\begin{equation}
\label{ns.eq:energyMomentum}
  T^{i}{}_{j} =
  \begin{pmatrix}[c]
    \rho + \f{q^{2}}{\kappa r^{4}}   & 0  & 0  & 0  \\
    0     & -p_{r}+ \f{q^{2}}{\kappa r^{4}}  & 0  & 0  \\
    0     & 0  & -\ppen- \f{q^{2}}{\kappa r^{4}} & 0  \\
    0     & 0  & 0  & -\ppen - \f{q^{2}}{\kappa r^{4}} 
  \end{pmatrix}.
\end{equation}
As a result, the complete Einstein-Maxwell system (EMS) of field
equations becomes
\begin{subequations}
  \label{ns.eq:EinMR}
  \begin{equation}
    \label{ns.eq:EinMR1}
    \kappa \rho +\f{q^{2}}{r^{4}} = \e^{-\lambda}\left( \f{\lambda'}{r} -\f{1}{r^2}\right) +\f{1}{r^{2}} 
    = \f{1}{r^{2}} - \f{Z}{r^{2}} - \f{1}{r}\deriv{Z}{r},
  \end{equation}
  \begin{equation}
  \label{ns.eq:EinMR2}
    \kappa p_{r} - \f{q^{2}}{r^{4}}= \e^{-\lambda} \left( \f{\nu'}{r} + \f{1}{r^2}\right) -\f{1}{r^2} 
    = \f{2Z}{rY}\deriv{Y}{r} + \f{Z}{r^{2}} - \f{1}{r^{2}},  
  \end{equation}
  \begin{multline}
    \label{ns.eq:EinMR3}
    \kappa \ppen + \f{q^{2}}{r^{4}} = \e^{-\lambda} \left( \f{\nu''}{2} - 
      \f{\nu'\lambda'}{4} + \f{(\nu')^2}{4} + \f{\nu'- \lambda'}{2r}\right)\\
    = \f{Z}{Y}\sderiv{Y}{r} + \f{1}{2Y}\deriv{Y}{r}\deriv{Z}{r} + \f{Z}{rY}\deriv{Y}{r} + \f{1}{2r}\deriv{Z}{r},
  \end{multline}
  \begin{equation}
    \label{ns:eq:EinMR4}
    F_{10} = -F_{01} = \f{q}{r^{2}}\e^{(\nu + \lambda)/2} = \f{qY}{r^{2}\sqrt{Z}},
  \end{equation}
\end{subequations}
The solution of this set of equations can be constructed in a variety
of ways, and for an extensive review, see Ref.~\cite{Iva02}.  However
the aim of this article is to find new solutions, that can be
considered to be physically relevant and therefore we proceed with a
method outlined in~\cite{Tol39}.

\subsection{Solution method}\label{ns.sec:solnMethod}
First we assume a functional form for the
\(Z\) metric function given by
\begin{equation}
\label{ns.eq:ZSol}
   Z(r) \eqqcolon 1- br^2+ ar^4.
\end{equation}
For the moment, \(a\) and \(b\) are undetermined constants, and \(r\)
is the radial coordinate.  The physical interpretation of the
constants will come later in Section~\ref{ns.sec:phys}.

This assumption helps in solving the system of
equation~\eqref{ns.eq:EinMR} for \(Z,\) since the first order
equation~\eqref{ns.eq:EinMR1} can immediately be integrated in terms
of one integration constant.  To find the solution for
\(Y,\) one first defines a new quantity,``the measure of anisotropy,''
\(\Delta\) through subtraction of~\eqref{ns.eq:EinMR3}
from~\eqref{ns.eq:EinMR2},
\begin{multline}
  \label{ns.eq:DeltaDef}
  \kappa \Delta = \kappa(p_{r} - \ppen) = \f{Z}{rY}\left( \deriv{Y}{r}\right)
  -\f{Z}{Y}\left( \sderiv{Y}{r} \right) -\f{1}{2Y}\left( \deriv{Z}{r}\right)\left( \deriv{Y}{r}\right) \\- \f{1}{2r}\left( \deriv{Z}{r}\right) + \f{Z}{r^{2}} - \f{1}{r^{2}} + \f{2q^{2}}{r^{4}}.
\end{multline}
This equation can be rearranged and simplified as a linear second
order ODE for \(Y,\) which can then be solved using a transformation
of the independent variable \(r.\)
\begin{multline}
  \label{ns.eq:YDiffR}
2r^{2}Z \left(\sderiv{Y}{r}\right)  + 
\left[ r^{2}\left(\deriv{Z}{r} \right) -2rZ \right] \deriv{Y}{r} + \\ +
\left( 2+2r^{2}\Delta -2Z +r \deriv{Z}{r} -\f{4q^{2}}{r^{2}}\right) Y = 0.   
\end{multline}
The second order ODE has both \(\Delta\) and \(q\) as undetermined
functions, which when set to zero transforms the ODE into that for the
Tolman~VII solution discussed in detail in~\cite{RagHob15}. Therefore
this procedure provides a generalization of the Tolman~VII solution.

The next step in the solution is the transformation to a new variable
\(x = r^{2}.\) A straight forward derivation then transforms the
derivative
\begin{subequations}
\[\deriv{}{r} \equiv
2\sqrt{x}\deriv{}{x},\] 
and similarly \[\sderiv{}{r} \equiv
4x\sderiv{}{x} + 2 \deriv{}{x}.\]  
\end{subequations}
Applying these to equation~\eqref{ns.eq:YDiffR} results in
\begin{multline*}
  2xZ \left( 4x \sderiv{Y}{x} + 2 \deriv{Y}{x}\right) + \left( 2
    \sqrt{x} \deriv{Y}{x} \right)
  \left( 2 x^{3/2} \deriv{Z}{x} - 2 \sqrt{x}Z\right) \\
  + \left[ 2 + 2x\Delta +\sqrt{x}\left(2\sqrt{x}\deriv{Z}{x}\right) -2Z  -\f{4q^{2}}{x}\right]Y = 0,
\end{multline*}
which can be rearranged into
\[8x^{2}Z \sderiv{Y}{x} + \left( \cancel{4xZ} + 4x^{2} \deriv{Z}{x}
    \cancel{-4xZ} \right) \deriv{Y}{x} + 2\left( 1+ x\Delta
    +x\deriv{Z}{x}-Z -\f{2q^{2}}{x} \right) Y = 0,\] or equivalently,
\begin{equation}
  \label{ns.eq:YDiffX}
  Z \sderiv{Y}{x} + \left(\f{1}{2} \deriv{Z}{x} \right) \deriv{Y}{x} + \left( \f{1+ x\Delta +x\deriv{Z}{x}-Z  - 2q^{2}/x} {4x^{2}} \right) Y = 0.
\end{equation}

The second step of the solution procedure involves another radial
variable change from \(x\) to \(\xi\) which is defined through
\begin{equation}
  \label{ns.eq:DefXi}
  \xi = \int_{0}^{x} \f{\d \bar{x}}{\sqrt{Z(\bar{x})}}  = \f{2}{\sqrt{a}} \acoth \left( \f{1+\sqrt{1 - bx + a x^{2}}}{\sqrt{a} x} \right).
\end{equation}
This induces a change in the \(x-\)derivatives, so that
\begin{subequations}
\label{ns.eq:XiDerivs}  
\begin{equation}
  \deriv{}{x} \equiv \f{1}{\sqrt{Z(x)}}\deriv{}{\xi}, \quad\quad \text{and,}  
  \end{equation}
  \begin{equation}
    \sderiv{}{x} \equiv \f{1}{Z} \sderiv{}{\xi} - \f{1}{2Z^{3/2}}
\deriv{Z}{x} \deriv{}{\xi}.
  \end{equation}
\end{subequations}
Applying these changes to the differential equation~\eqref{ns.eq:YDiffX}
results in the elimination of the first derivative term for \(Y,\)
further simplifying the second order ODE:
\begin{multline}
\label{eq:Zsimp}
Z \left\{ \f{1}{Z} \sderiv{Y}{\xi} - \cancel{ \f{1}{2Z^{3/2}} \deriv{Z}{x} \deriv{Y}{\xi}}  \right\} 
+\cancel{ \f{1}{2} \deriv{Z}{x} \left( \f{1}{\sqrt{Z}} \deriv{Y}{\xi}\right) } \\
+ \left( \f{1+ x\Delta +x\deriv{Z}{x}-Z-2q^{2}/x}{4x^{2}} \right) Y = 0.  
\end{multline}
Since \(Z\)
is a known function of \(x,\)
simplification of the coefficient multiplying \(Y,\)
is immediately possible.  This reduces equation~\eqref{eq:Zsimp} into
a simple linear ODE since, 
\[ \left( \f{1+x\Delta+x(2ax-b) - (1-bx+ax^{2}) - 2q^{2}/x}{4x^{2}}
  \right) = \left( \f{a}{4} + \f{\Delta}{4x} -\f{q^{2}}{2x^{3}}
  \right),\] so that the ODE for \(Y\) finally becomes
\begin{equation}
  \label{ns.eq:YdiffXi2}
  \sderiv{Y}{\xi} + \left( \f{a}{4} + \f{\Delta}{4x} -\f{q^{2}}{2x^{3}} \right)Y = 0.
\end{equation}
When the coefficient multiplying \(Y\) is a constant, simple closed
form solutions are easily obtained.  As mentioned previously,
\(\Delta\) is a function that can be arbitrarily chosen, and is
interpreted as a measure of anisotropy between the pressures in the
solution.  From spherical symmetry, both the radial pressure \(p_{r}\)
and the tangential pressure \(\ppen\) must be equal at the centre,
resulting in \(\Delta\) having to be equal to zero when \(x=r^{2}=0.\)
Similarly the enclosed charge \(q\) can be given arbitrarily.  In
order to be physically relevant \(q\) must be regular everywhere
including at the origin where \(r=0.\)

\subsection{Physical considerations}\label{ns.sec:phys}
Following~\cite{RagHob15}, where a straightforward interpretation of
the Tolman~VII solution in terms of physical parameters of the star
was given, a physical motivation for the different functions in the
equation~\eqref{ns.eq:YdiffXi2} above is sought.  This section can be
seen as the motivation for choosing the radial dependence for
\(\Delta\) and \(q.\)

Considering that in the Tolman~VII solution, the ansatz for \(Z\)
allows the integration of equation~\eqref{ns.eq:EinMR1} directly,
proceeding in a similar fashion here simplifies the
ODE~\eqref{ns.eq:EinMR1}, with the caveat that now there is a
contribution from the charge \(q.\)

However, in general relativity, the electric charge has mass--energy,
and following Ivanov~\cite{Iva02} the external perceived mass is
redefined as the sum of the material rest mass, and the electrostatic
energy contained in the electrostatic field, so that
\begin{equation}
  \label{ns.eq:Mass+Charge}
  M = 4 \pi \int_{0}^{r_{b}} \left(\rho(r) +\f{q^{2}(r)}{8 \pi r^{4}} \right) r^{2} \d r.
\end{equation}
This redefinition of mass then results in the integration of
equation~\eqref{ns.eq:EinMR1} as follows, since \(Z\)
has already an assumed form,
\begin{equation}
  \label{ns.eq:EinMR1Int}
M = 4 \pi \int_{0}^{r_{b}} \left( 1 - Z - r \deriv{Z}{r} \right) \d r =  
4 \pi \int_{0}^{r_{b}} (3br^{2} - 5 a r^{4}) \d r.
\end{equation}
This mass equation leads to a physical interpretation for the
constants \(a\) and \(b.\) Since the Tolman~VII solution has physical
characteristics, \emph{and} is compatible with the form of \(Z\)
chosen, one assumes the same form in all of the future solutions.  The
physical reasoning motivating this choice is given in detail
in~\cite{RagHob15}, where a quadratic density profile of the form
\begin{equation}
\label{ns.eq:ZSol1}
    \rho = \rho_{c} \left[ 1 - \mu \left( \f{r}{r_{b}}\right)^{2}
\right],
  \end{equation}
  leads to a metric function \(Z\) becomes
\begin{equation}
\label{ns.eq:ZSol2}
 Z(r)= 1 - \left( \f{\kappa
     \rho_{c}}{3}\right) r^{2} + \left(\f{\kappa \mu
     \rho_{c}}{5r_{b}^{2}}\right) r^{4} = 1- br^2+ ar^4.
\end{equation}
Thus \(a\) and \(b\) can be identified with the parameters in
parentheses.  In~\eqref{ns.eq:ZSol1}, the constant \( r_{b} \)
represents the boundary radius at the matter-vacuum interface, and
\( \rho_{c} \) represents the central density at \( r=0.\) Finally
\( \mu \) is a dimensionless ``self-boundedness'' parameter where
\(0 \leq \mu \leq 1.\) When \(\mu=0\), one has a sphere of constant
density, and when \(\mu=1,\) a ``natural'' star, with density
vanishing at the boundary is obtained.  For intermediate values there
is a density discontinuity at \(r_{b}\) which decreases in magnitude
with increasing \(\mu.\)

This same property shall be used once again in
Section~\ref{ns.sec:Ani}, where an anisotropic pressure only is
considered.  For the case with charge a more general approach will be
taken.

The charge distribution will be assumed to take the form
\(q(r) = kr^{n},\) where \(k\) is a constant and \(n>0.\) Such charge
distributions occur in many static situations, for mathematical
reasons (see~\cite{Iva02} for example) but they can be expected to
exist on physical grounds as well.

First all known charge carriers are massive.  Therefore they will also
undergo gravitational attraction.  In addition the static
electromagnetic field energy density is positive since it is
proportional to \(q^{2}\) which then also contributes to the
attractive gravitational force.  On the other hand it is known that in
the absence of any other forces all free charges reside on the surface
of a conducting medium.

It can therefore be expected that when gravitational and
electromagnetic forces are closed to being balanced, the strength of
the electromagnetic repulsion will tend to concentrate charge in the
outer regions of a compact fluid body, while maintaining a non-zero
charge density inside the body.

The fact that pressure terms are introduced due to non-gravitational,
non-electromagnetic interactions helps provide more realistic charge
distributions than are found in charged dust solutions for example.

From the above discussion the initial ansatz for \(Z\)
from equation~\eqref{ns.eq:ZSol} can be substituted into the RHS of the
first ODE~\eqref{ns.eq:EinMR1}, which results in
\begin{equation}
\label{ns.eq:qZ}
\kappa \rho + \f{q^{2}}{r^{4}} = 3b - 5ar^{2}.  
\end{equation}
Consistency, and the desire to keep the procedure for solving this
system of equations the same as outlined in
Section~\ref{ns.sec:solnMethod} then demands that the LHS of the
equation~\eqref{ns.eq:qZ} also be a quadratic function with vanishing
linear term.  Because of the structure of this equation one is forced
to choose either \(q(r) = k r^{2},\) in which case,
\[ 3b - 5ar^{2} = (\kappa \rho_{c} + k^{2}) -
\f{\kappa\rho_{c}\mu}{r_{b}^{2}} r^{2},\]
or  \(q(r) = k r^{3},\)
which results in
\[ 3b - 5ar^{2} = \kappa \rho_{c} - \left(
    \f{\kappa\rho_{c}\mu}{r_{b}^{2}} - k^{2}\right) r^{2}.\] 

Therefore the charge introduces terms that redefine the parameters
\(a\) and \(b,\) in terms of physical parameters
\(\rho_{c}, \mu, r_{b}\) and \(k.\) However, it is found that the
choice \(q(r) = k r^{2}\) yields a differential equation for \(Y(r)\)
that is not soluble with elementary functions\footnote{The coefficient
  of \(Y\) in the second order ODE after variable changes still
  contains a $1/r^2$ term, turning the problem into a variable
  coefficient one. Once additional assumptions about $a$ have been
  made, a solution in terms of hypergeometric functions is possible,
  but the assumption about $a$ renders the solution physically
  uninteresting.}.  Discarding the quadratic radial dependence, the
choice \(q(r) =kr^{3}\) is selected for the remainder of this article.

Turning to the ``measure of anisotropy'' \(\Delta(r),\) one has to
satisfy the condition that the metric is spherically symmetric.  Both
pressures \emph{must} be equal at the coordinate centre, \(r=0\) if
spherical symmetry is to be satisfied, and this can be translated into
\(\Delta(0) = 0.\) Except for this condition, there is complete
arbitrariness in specifying the anisotropy, and indeed the literature
is replete with different ans\"atze for \(\Delta.\) For the purely
mathematical reason of solving equation~\eqref{ns.eq:YdiffXi2} the
choice we make is \(\Delta = \beta x = \beta r^{2}.\) This choice
satisfies the condition at the centre, \emph{and} leads to a constant
value for the coefficient of \(Y\) appearing
equation~\eqref{ns.eq:YdiffXi2}.

With these two assumptions about the anisotropy and charge, the ODE
for \(Y\) becomes
\begin{equation}
  \label{ns.eq:YdiffXi}
  \sderiv{Y}{\xi} = - \f{Y}{4}\left( a + \beta - 2k^{2} \right) = -\Phi^{2} Y.
\end{equation}

\begin{figure}[p]
  \includegraphics[angle=90, width=\textwidth]{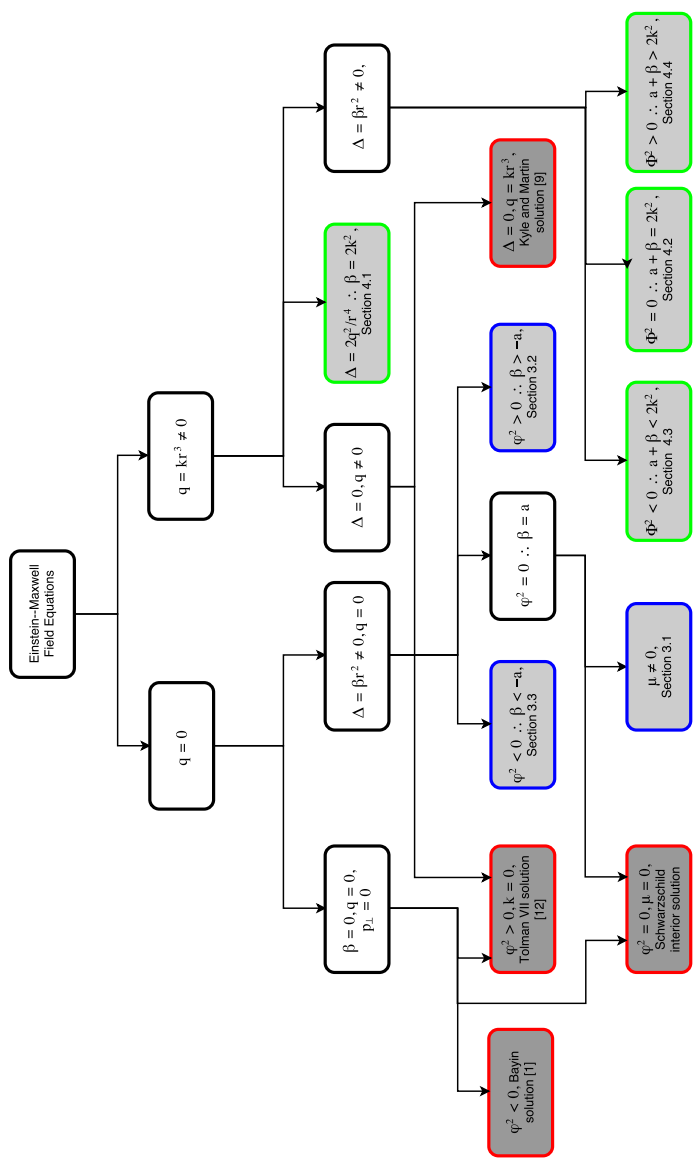}
  \caption{The solution landscape explored in this article. Lightly
    shaded boxes are the new solutions described in this work, and
    darker ones are the older known solutions.  In the online coloured
    version, red-bordered boxes are solutions with isotropic pressure,
    blue-bordered ones are uncharged solutions with anisotropic
    pressures only and green-bordered ones are charged with
    anisotropic pressure.}
\label{ns.fig:SolFlow}
\end{figure}

Different solutions can be generated from different assumptions about
\(\Phi^{2}, \beta\) or \(k,\) and a classification that summarizes
these choices is provided in Figure~\ref{ns.fig:SolFlow}.
Those in the lightly shaded boxes are the ones that will be discussed
in this article.

\subsection{Boundary conditions}
The constants of integration associated with each solution will be
determined by the boundary conditions (BCs) that one specifies.  The
system of equations consists of one first order ODE for \(Z,\) and one
second order ODE for \(Y.\) The first one makes use of the definition
of mass in equation~\eqref{ns.eq:Mass+Charge}.  For the second one, we
require two BCs, and the junction conditions matching the interior
metric to the exterior metric provide the two needed BCs.

These junction conditions in the case of spherically symmetric and
static solutions, match the interior solution to the charged exterior
Reissner--Nordstr\"om metric given by
\begin{equation}
  \label{ns.eq:Reiss}
  \d s^{2} = \left( 1 - \f{2M}{r} + \f{Q^{2}}{r^{2}}\right) \d t^{2} - 
\left( 1 - \f{2M}{r} + \f{Q^{2}}{r^{2}}\right)^{-1} \d r^{2} - r^{2}\left( \d \theta^{2} + \sin^{2} \theta \d \varphi^{2} \right),
\end{equation}
where \(M\) is the mass given by~\eqref{ns.eq:Mass+Charge} and \(Q\) the charge given by
\[Q = q(r_{b}).\]  
\begin{subequations} 
The radial pressure which can be computed once the solution for
  \(Y\)
  has been specified, has to vanish at the boundary of the compact
  object where the exterior and interior metric meet. Therefore
\label{ns.eq:Boundary1+2}
  \begin{align}
    \label{ns.eq:BoundaryPca}
    p_{r}(r_{b}) &= 0, \quad \text{and,} \\
    \label{ns.eq:BoundaryZca}
    Z(r_{b}) &= 1-\f{2M}{r_{b}} + \f{Q^{2}}{r^{2}_{b}} = Y^{2}(r_{b}).
  \end{align}
\end{subequations}
The second boundary condition~\eqref{ns.eq:BoundaryZca} can also be
implemented once the solution for \(Y\)
has been specified.  It is possible to devise an expression for the
radial pressure \(p_{r}\)
solely in terms of the metric and metric derivatives that make
applying the first boundary condition easier.  The first step in
generating this expression is the addition of
equations~\eqref{ns.eq:EinMR1} and~\eqref{ns.eq:EinMR1} to obtain
\begin{equation}
  \label{ns.eq:EinR1+2}
  \kappa (p_{r}+\rho) = \e^{-\lambda} \left( \f{\nu'}{r} + \f{\lambda'}{r} \right) = \f{2Z}{rY}\deriv{Y}{r} - \f{1}{r}\deriv{Z}{r}.
\end{equation}
Going through the same coordinate changes as discussed in
Section~\ref{ns.sec:solnMethod} one obtains
\[\kappa p_{r} = \f{2Z}{rY}\deriv{Y}{r} - \f{1}{r}\deriv{Z}{r} -\kappa \rho \xrightarrow{r
  \rightarrow x} \f{4Z}{Y}\deriv{Y}{x} - 2\deriv{Z}{x} -\kappa \rho
\xrightarrow{x \rightarrow \xi} \f{4\sqrt{Z}}{Y} \deriv{Y}{\xi} - 2
\deriv{Z}{x} -\kappa \rho. \]
This expression can be evaluated at the boundary \(r=r_{b}\),
with conditions~\eqref{ns.eq:Boundary1+2} simplifying the results,
\[ \kappa p_{r}(r_{b}) = 0 =
\f{4\cancel{\sqrt{Z(r_{b})}}}{\cancel{Y(r_{b})}}
\left. \deriv{Y}{\xi}\right|_{\xi = \xi_{b}} - \left. 2
  \deriv{Z}{x}\right|_{x=x_{b}} -\kappa \rho(r_{b}), \]
so that 
\[
  \kappa\rho(r_{b}) = \left. 4\deriv{Y}{\xi}\right|_{\xi=\xi_{b}} - \left. 2\deriv{Z}{x}\right|_{x=x_{b}}.
\]
This leads to a general boundary condition on the derivative of \(Y\)
\begin{equation}
    \label{ns.eq:BoundaryApp1}
\left. \deriv{Y}{\xi}\right|_{\xi=\xi_{b}} = \f{1}{4} \left[ \f{\kappa \rho_{c}}{3} - \f{\kappa\rho_{c}\mu}{5} -\f{4k^{2}r_{b}^{2}}{5}\right] =: \alpha.
\end{equation}
For the second condition one re-expresses equation~\eqref{ns.eq:BoundaryZca} in terms of \(Y\) through
\begin{equation}  \label{ns.eq:BoundaryApp2}
Y(r_{b}) = \sqrt{Z(r_{b})} = \sqrt{1 + \f{\kappa \rho_{c}r_{b}^{2}(3\mu-5)}{15} - \f{k^{2}r_{b}^{4}}{5}} =: \gamma.
\end{equation}
Subsequent application of these two conditions on the \(Y\) metric
function form a Cauchy boundary pair and results in unique integration
constants for the \(Y\) metric function in terms of the auxiliary
constants \(\alpha\) and \(\gamma,\) defined in
equations~\eqref{ns.eq:BoundaryApp1} and~\eqref{ns.eq:BoundaryApp2}
respectively.

\subsection{New Solutions}
The spherically symmetric Einstein--Maxwell interior field equations
can now be solved completely since a distinct ODE whose solutions are
available has been presented, and furthermore, one can interpret all
the constants present in the solutions physically.  The availability
of boundary conditions ensures the uniqueness of the solution.  Based
on Figure~\ref{ns.fig:SolFlow} the ODE for \(Y\) for different
physical conditions will be discussed in the following sections.

\section{Uncharged case with anisotropic pressures}\label{ns.sec:Ani}
In this section, only the uncharged \(q = k = 0\) specialization of
the ODE~\eqref{ns.eq:YdiffXi} is investigated.  As a result, this
equation becomes
\[\sderiv{Y}{\xi} + \left( \f{a+\beta}{4}\right)Y = 0.\]  
The solutions can immediately be written in terms of the parameter
\(\phi^{2} = \Phi^{2}(k=0) = (a+\beta)/4\)
in the following table.  The details of the properties associated with
the different solutions are discussed in the sections indicated in
Table~\ref{ns.tab:AniSolns}.
\begin{table}[!h]
\centering 
\begin{tabular}{ >{$}c<{$}  >{$}c<{$}  c }
    \phi^2 & Y(\xi) & Solution's analysis \\
    \hline\hline
    \phi^2 < 0 & c_1 \cosh{\left(\sqrt{-\phi^2}\xi\right)} + c_2 \sinh{\left(-\sqrt{-\phi^2}\xi\right)} & section~\ref{ns.ssec:phiNeg} \\
    \phi^2 = 0 & c_1 + c_2 \xi & section~\ref{ns.ssec:phiZero}\\
    \phi^2 > 0 & c_1 \cos{\left( \phi \xi \right)} + c_2 \sin{\left(\phi \xi\right)} & section~\ref{ns.ssec:phiPos}\\
  \end{tabular}
  \caption[The different solutions with $\phi=0$]{The different solutions that can be generated through different values of the parameter~$\phi^{2}.$ The integration constants $c_{1},$ and $c_{2}$ are determined by the two boundary conditions.}
\label{ns.tab:AniSolns}
\end{table}

\subsection{The $ \phi^{2} = 0 $ case}\label{ns.ssec:phiZero}
When \(\phi^{2} = 0,\) the only possibility is for
\(\beta = -a = -\f{\kappa\mu\rho_{c}}{5r_{b}^{2}},\) which is either
negative when all the parameters in the
\(\f{\kappa\mu\rho_{c}}{5r_{b}^{2}}\) expression are positive
definite: the case we will consider now, or zero when the parameter
\(\mu = 0.\) The latter case reduces to the Schwarzschild interior
solution for which there is much historical~\cite{Tol66,Wey52} and
contemporary literature~\cite{Wal84,Inv92}.  For the \(\beta \neq 0\)
case, \(\ppen = p_{r} - \Delta = p_{r} + ax,\) and the angular
pressure is thus larger than the radial pressure everywhere but at the
centre.  Applying the two boundary conditions to solve for the
integration constants, equation~\eqref{ns.eq:BoundaryApp1} yields

\begin{equation}
\label{eq:Cauchy1}
\left. \deriv{Y}{\xi} \right|_{\xi=\xi_{b}} = c_{2} = \alpha \coloneqq \f{1}{4} \left( \f{\kappa \rho_{c}}{3} - \f{\mu \kappa \rho_{c}}{5} \right),  
\end{equation}
where the \(\xi-\)derivative can be computed for \(Y\) from its
expression given in Table~\ref{ns.tab:AniSolns}.  The constant
\(\alpha\) can be obtained from expressions for the density and \(Z\)
given in equation~\eqref{ns.eq:ZSol1} and~\eqref{ns.eq:ZSol2}.

Using equation~\eqref{ns.eq:BoundaryApp2}
\begin{equation}
\label{eq:Cauchy2}
\left. Y \right|_{\xi=\xi_{b}} = c_{1} +
c_{2}\xi_{b} = \gamma \coloneqq \sqrt{Z(r_{b})} = \sqrt{1 - \left(
    \f{\kappa \rho_{c}}{3}\right) r_{b}^{2} + \left(\f{\kappa \mu
      \rho_{c}}{5r_{b}^{2}}\right) r_{b}^{4}},  
\end{equation}
it can be deduced that
\[c_{1} = \gamma - \f{2 \alpha}{\sqrt{a}} \acoth{\left(
    \f{1+\gamma}{r^{2}_{b} \sqrt{a}} \right)}. \]

\begin{figure}[h!]
  \centering
  \includegraphics[width=\linewidth]{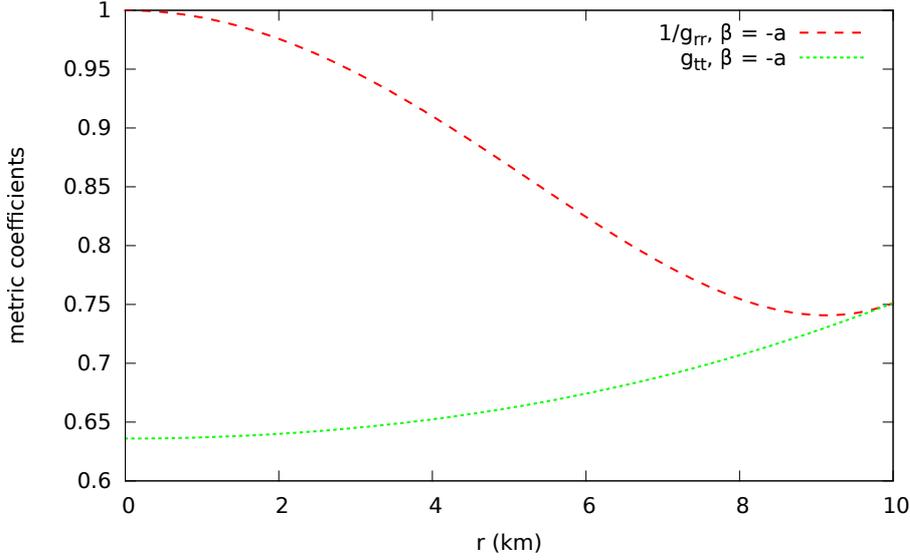}
  \caption[Matching the metric functions for $\phi^{2} =0$ (Anisotropy
  only)]{Application of the boundary conditions that match the value
    and radial derivative of the metric function at $r=r_{b}$ for the
    $\phi^{2} =0$ case. The parameter values are
    $\rho_{c}=\un{1\times 10^{18}\,kg \cdot m^{-3}}, r_b = \un{1
      \times 10^{4}\,m}$ and are chosen to give the conditions that
    might exist for a physically acceptable compact star. The
    parameter $\mu = 1$ is chosen for simplicity.}
\label{ns.fig:MetricPhiZ}
\end{figure}

Figure~\ref{ns.fig:MetricPhiZ} plots the metric functions and shows
that the metric functions and their derivatives become equal at the
radius \(r_{b},\)
as expected from the matching to the Schwarzschild exterior metric, at
the boundary \(r_{b}.\)

The solution for the second metric function as a function of \(r\)
can, after the changes in the radial parameter, be given as
\begin{equation}
  \label{ns.eq:PhiZY}
  Y(r) = \gamma + \f{2\alpha r_{b}}{\sqrt{\kappa \rho_{c}\mu /5}}\left[ 
  \acoth{\left(\f{1-\sqrt{Z(r)}}{r^{2}\sqrt{\f{\kappa \rho_{c} \mu} {5r_{b}^{2}}}}\right) }  -\acoth{\left( \f{1-\gamma}{r_{b}\sqrt{\kappa \rho_{c} \mu /5}}\right)}  \right],
\end{equation}
where the constants \(\alpha\) and \(\gamma\) are given in terms of
the fundamental set of parameters \(r_{b}, \rho_{c}\) and \(\mu\) by equations~\eqref{eq:Cauchy1} and~\eqref{eq:Cauchy2}.

The two pressures can similarly be given in terms of the above
variables.  The radial pressure can be computed from the second
Einstein equation~\eqref{ns.eq:EinMR2} in a straightforward manner to yield
\begin{multline}
  \label{ns.eq:PhiZpr}
  \kappa p_{r}(r) = \f{2\kappa\rho_{c}}{3} - \f{4\kappa\rho_{c}\mu r^{2}}{5r_{b}^{2}} -\kappa\rho_{c} \left[ 1 - \mu \left( \f{r}{r_{b}}\right)^{2}\right] + 
 \left( \f{\kappa\rho_{c}}{3} - \f{\kappa \rho_{c} \mu}{5}\right) \times ... \\ \times \f{\sqrt{1-\f{\kappa\rho_{c}}{3}r^{2} + \f{\kappa\mu\rho_{c}}{5r_{b}^{2}} r^{4}}}{\gamma + \f{2\alpha r_{b}}{\sqrt{\kappa \rho_{c}\mu /5}}\left[ 
  \acoth{\left(\f{1-\sqrt{Z(r)}}{r^{2}\sqrt{\f{\kappa \rho_{c} \mu} {5r_{b}^{2}}}}\right) }  -\acoth{\left( \f{1-\gamma}{r_{b}\sqrt{\kappa \rho_{c} \mu /5}}\right)}  \right]},
\end{multline}
and similarly the tangential pressure is easily written in terms of the above as
\begin{equation}
  \label{ns.eq:PhiZpt}
  \ppen(r) =p_{r} - \beta r^{2} = p_{r} + \f{\kappa\rho_{c}\mu}{5r_{b}^{2}} r^{2}. 
\end{equation}
This completes the solution, since all the metric functions and matter
variables have been found in terms of the parameters in the ansatz and
the radial coordinate \(r\).  If an equation of state for this
solution is required, one could invert the density
relation~\eqref{ns.eq:ZSol1}, to get an expression for \(r\) in terms
of \(\rho,\) such that \(r = r_{b}\sqrt{(1-\rho/\rho_{c})/\mu}.\)
Simple substitution in the expressions for the
pressures~\eqref{ns.eq:PhiZpr} and~\eqref{ns.eq:PhiZpt} will then give
the equation of state for both pressures \(p_{t}(\rho),\) and
\(\ppen(\rho),\) in a process similar to what was done in
Ref~\cite{RagHob15}.

\subsection{The $ \phi^{2} > 0 $ case}\label{ns.ssec:phiPos}
When \(\phi^{2} > 0, a + \beta > 0,\) which implies that
\(\beta > -a.\) Since \(a\) has a specific dependence on the
parameters \(\rho_{c}, \mu,\) and \(r_{b},\) one obtains
\(\beta > - \f{\kappa \mu \rho_{c}}{5r_{b}^{2}}, \) which allows
\(\beta\) to have negative values, since the fraction in the last
expression is positive definite.  \(Y\) is then given in terms of the
trigonometric functions given in Table~\ref{ns.tab:AniSolns}, from
which one can write expressions for the derivative of
\(\deriv{Y}{\xi}\) by direct computation. Applying this boundary
condition leads to the following relation among the constants of
integration and the free parameters that describe the solution:
\[\left. \deriv{Y}{\xi} \right|_{\xi=\xi_{b}} =  \phi \left[ c_{2} \cos{(\phi \xi_{b})} - c_{1} \sin{(\phi \xi_{b})} \right] =\alpha \coloneqq  \f{\kappa\rho_{c}\left( 5-3\mu\right)}{60}.\] 

\begin{figure}[h!]
  \centering
  \includegraphics[width=\linewidth]{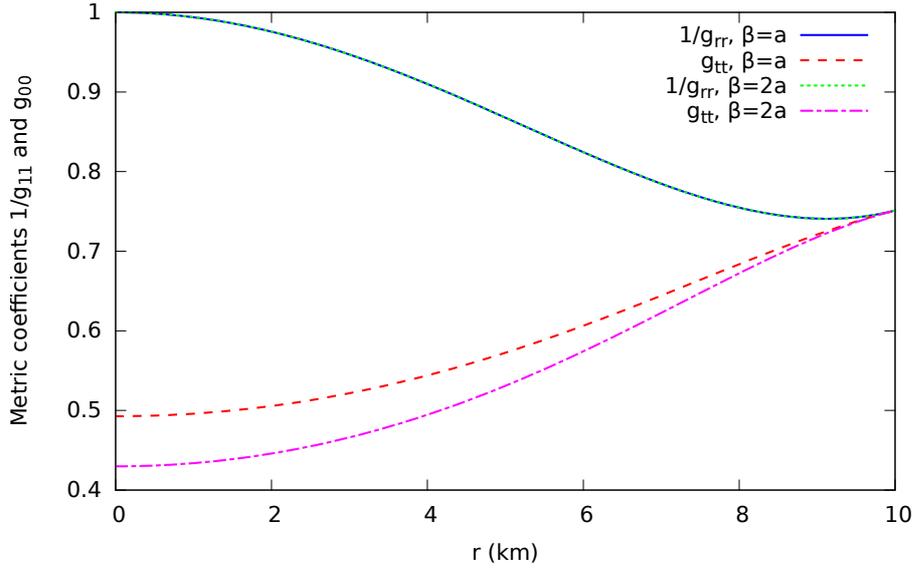}
  \caption[Matching the metric functions for $\phi > 0$ (Anisotropy
  only)]{Application of the boundary conditions that match the value
    and radial derivative of the metric function at $r=r_{b}$ for the
    $\phi^{2} > 0$ case.  The parameter values are
    $\rho_{c}=\un{1\times 10^{18}\,kg \cdot m^{-3}}, r_b = \un{1
      \times 10^{4}\,m}$ and $\mu = 1,$ with $\beta$ given in the
    legend. Again the parameters are chosen such that the solution is
    applicable to a model of a compact star.}
\label{ns.fig:MetricPhiP}
\end{figure}

Furthermore the boundary condition applied to \(Y\) leads to:
\[\left. Y \right|_{\xi=\xi_{b}} = c_{2} \sin{(\phi \xi_{b})} +
  c_{1}\cos{(\phi \xi_{b})} = \gamma \coloneqq \sqrt{1+ \f{\kappa \rho_{c}r_{b}^{2}(3\mu-5)}{15} }.\] 
These two equations provide a means for solving for \(c_{1}\) and \(c_{2},\) to obtain
\begin{align*}
  c_{2} &= \gamma \sin{(\phi\xi_{b})} + \f{\alpha}{\phi} \cos{(\phi\xi_{b})}\\
  c_{1} &= \gamma \cos{(\phi\xi_{b})} - \f{\alpha}{\phi} \sin{(\phi\xi_{b})}.
\end{align*}
A plot of the metric functions is shown in
Figure~\ref{ns.fig:MetricPhiP} and shows how the metric functions and
their radial derivatives match at the boundary \(r=r_{b}.\)
The complete solution for the \(Y-\)metric function in this case is
  \begin{multline}
\label{ns.eq:YphiP}
    Y(r) = \\
\left[ \gamma \cos{(\phi\xi_{b})} - \f{\alpha}{\phi} \sin{(\phi\xi_{b})} \right] 
    \cos{\left[ \f{2\phi}{\sqrt{a}} \coth^{-1} \left( \f{1+\sqrt{1-br^2+ar^4}}{r^2\sqrt{a}}\right)  \right]} +\\
    + \left[\gamma \sin{(\phi\xi_{b})} + \f{\alpha}{\phi} \cos{(\phi\xi_{b})} \right] 
    \sin{\left[\f{2\phi}{\sqrt{a}} \coth^{-1} \left( \f{1+\sqrt{1-br^2+ar^4}}{r^2\sqrt{a}}\right) \right]},
  \end{multline}
which then leads to explicit expressions for the pressures \(\ppen\) and \(p_{r}\) as
\begin{multline}
  \label{ns.eq:PhiPpr}
\kappa p_{r}(r) = \f{2\kappa\rho_{c}}{3} - \f{4\kappa\rho_{c}\mu r^{2}}{5r_{b}^{2}} -\kappa\rho_{c} \left[ 1 - \mu \left( \f{r}{r_{b}}\right)^{2}\right] + 4 \phi\sqrt{1-br^{2}+ar^{4}} \times \\
\times \f{\left[ \gamma \sin{(\phi\xi_{b})} + \f{\alpha}{\phi} \cos{(\phi\xi_{b})} \right] \cos{(\phi \xi)} - \left[ \gamma \cos{(\phi\xi_{b})} - \f{\alpha}{\phi} \sin{(\phi\xi_{b})} \right] \sin{(\phi \xi)} }{\left[ \gamma \sin{(\phi\xi_{b})} + \f{\alpha}{\phi} \cos{(\phi\xi_{b})}\right] \sin{(\phi \xi)} + \left[\gamma \cos{(\phi\xi_{b})} - \f{\alpha}{\phi} \sin{(\phi\xi_{b})} \right]\cos{(\phi \xi)}},  
\end{multline}
and 
\begin{equation}
  \label{ns.eq:PhiPpt}
  \ppen(r) =p_{r} - \beta r^{2}. 
\end{equation}
As in the previous examples, and in particular the Tolman~VII
solution, inversion of the density--radial coordinate relation
generates an equation of state, if required.

\subsection{The $ \phi^{2} < 0 $ case}\label{ns.ssec:phiNeg} 
When \(\phi^{2} < 0, a + \beta < 0,\)
which implies that \(\beta < -a.\)
Again using the explicit expression for
\(a, \beta < - \f{\kappa \mu \rho_{c}}{5r_{b}^{2}}, \)
which forces \(\beta\)
to have negative values, since the fraction in the last expression is
positive definite.  The expression for the derivative of \(Y\)
can be found by direct computation.  Applying the boundary
condition on \(\deriv{Y}{\xi}\) at \(\xi_{b}\) leads to,
\[\left. \deriv{Y}{\xi} \right|_{\xi=\xi_{b}} =  \phi \left[ c_{2} \cosh{(\phi \xi_{b})} + c_{1} \sinh{(\phi \xi_{b})} \right] =\alpha,\] 
while a similar calculation for \(Y\) at \(\xi_{b}\) yields
\[\left. Y \right|_{\xi=\xi_{b}} = \gamma \Rightarrow c_{2} \sinh{(\phi \xi_{b})} +
  c_{1}\cosh{(\phi \xi_{b})} = \gamma.\]
These two equations provide explicit expressions for the integration constants,
\begin{align*}
  c_{2} &= \f{\alpha}{\phi} \cosh{(\phi\xi_{b})} - \gamma \sinh{(\phi\xi_{b})}  \\
  c_{1} &= \gamma \cosh{(\phi\xi_{b})} - \f{\alpha}{\phi} \sinh{(\phi\xi_{b})},
\end{align*}
where \(\alpha\) and \(\gamma\) have the same expressions given in
equation~\eqref{eq:Cauchy1} and~\eqref{eq:Cauchy2}.
 
\begin{figure}[h!]
  \centering
  \includegraphics[width=\linewidth]{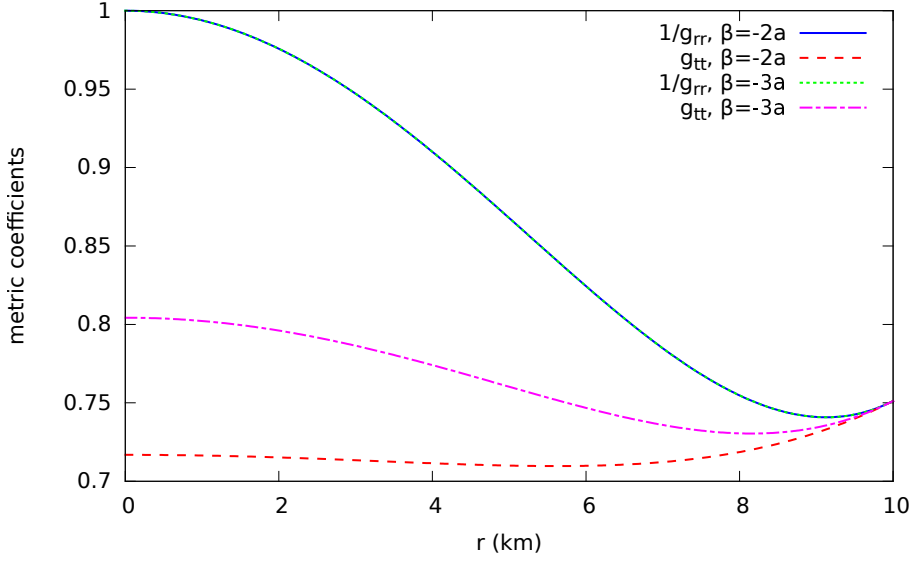}
  \caption[Matching the metric functions for $\phi < 0$ (Anisotropy
  only)]{Application of the boundary conditions that match the value
    and radial derivative of the metric function at $r=r_{b}.$ for the
    $\phi < 0$ case.  The parameter values are
    $\rho_{c}=\un{1\times 10^{18}\,kg \cdot m^{-3}}, r_b = \un{1
      \times 10^{4}\,m}$ and $\mu = 1,$ with $\beta$ given in the
    legend.  Again the values for the constants have been chosen to
    generate models of stars that could be used for compact objects.}
\label{ns.fig:MetricPhiN}
\end{figure}

Plotting the metric functions as a function of \(r,\)
Figure~\ref{ns.fig:MetricPhiN} again shows the result of matching the
metric functions and their derivatives to exterior Schwarzschild
solution at the boundary \(r=r_{b}.\)

The complete solution for the \(Y-\)metric function in this case is

\begin{multline}
\label{ns.eq:YphiN}
    Y(r) = \\
    \left[ \gamma \cosh{(\phi\xi_{b})} - \f{\alpha}{\phi} \sinh{(\phi\xi_{b})} \right] 
    \cosh{\left[ \f{2\phi}{\sqrt{a}} \acoth \left( \f{1+\sqrt{1-br^2+ar^4}}{r^2\sqrt{a}}\right)  \right]} +\\
    + \left[ \f{\alpha}{\phi} \cosh{(\phi\xi_{b})} - \gamma \sinh{(\phi\xi_{b})} \right] 
    \sinh{\left[\f{2\phi}{\sqrt{a}} \acoth \left( \f{1+\sqrt{1-br^2+ar^4}}{r^2\sqrt{a}}\right) \right]},
\end{multline}
which then allows the matter variable \(p_{r}\) to be written as
\begin{multline}
  \label{ns.eq:PhiNpr}
\kappa p_{r}(r) = \f{2\kappa\rho_{c}}{3} - \f{4\kappa\rho_{c}\mu r^{2}}{5r_{b}^{2}} -\kappa\rho_{c} \left[ 1 - \mu \left( \f{r}{r_{b}}\right)^{2}\right] + 4 \phi\sqrt{1-br^{2}+ar^{4}} \times \\
\times \f{ \left[ \f{\alpha}{\phi} \cosh{(\phi\xi_{b})} -\gamma \sinh{(\phi\xi_{b})} \right] \cosh{(\phi \xi)} + \left[ \gamma \cosh{(\phi\xi_{b})} - \f{\alpha}{\phi} \sinh{(\phi\xi_{b})} \right] \sinh{(\phi \xi)} }{\left[ \gamma \cosh{(\phi\xi_{b})} - \f{\alpha}{\phi} \sinh{(\phi\xi_{b})}\right] \cosh{(\phi \xi)} + \left[\gamma \cosh{(\phi\xi_{b})} - \f{\alpha}{\phi} \sinh{(\phi\xi_{b})} \right]\sinh{(\phi \xi)}},  
\end{multline}
and \(\ppen,\) the tangential pressure is expressed through the above as
\begin{equation}
  \label{ns.eq:PhiNpt}
  \ppen(r) =p_{r} - \beta r^{2}. 
\end{equation}

As with the previous examples, an explicit equation of state can be
obtained from inverting equation~\eqref{ns.eq:ZSol1} to obtain \(r(\rho).\)

In the next section charged generalizations of solutions with
anisotropic pressures will be discussed.

\section{Charged case with anisotropic pressures}\label{ns.sec:Cha}
In this section electrically charged solutions with
\(q =kr^{3} \neq 0\) are investigated.  The simplest case sets
\(\beta = 0,\) which eliminates the anisotropic pressure.  This
results in \[4\Phi^{2} = a - 2k^{2}.\] Setting \(k=0\) leads to the
Tolman~VII solution as expected.  The general case with \(k\neq 0\)
yields the solution given by Kyle and Martin~\cite{KylMar67}.

Now consider the \(\Delta \neq 0, \) case instead. Requiring that
\(\f{\Delta}{x} = \f{2q^{2}}{x^{3}}\) effectively ``anisotropises''
the electric charge allowing the latter to contribute to the
anisotropy only, and considerably simplifies the solution to \(Y.\)
This solution is examined in Section~\ref{ns.ssec:phiA}.  If instead
\(\Delta = \beta x,\) and \(2q^{2} = 2 k^{2} x^{3},\) one gets
\(4\Phi^{2} = a + \beta -2k^{2},\) which allows an analysis very
similar to what was done in the previous Sections since \(\Phi^{2}\)
can then be of either sign.  This possibility is investigated in
Sections~\ref{ns.sssec:phi0} to ~\ref{ns.sssec:phiP}

\subsection{``Anisotropised charge''}\label{ns.ssec:phiA}
\begin{figure}[h!]
  \centering
  \includegraphics[width=\linewidth]{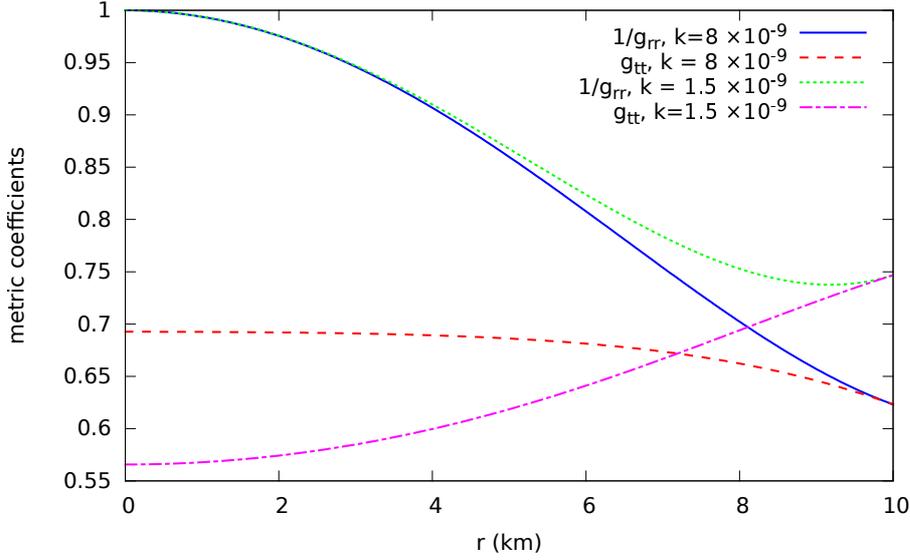}
  \caption[Matching the metric functions, for the anisotropy
  compensating the charge]{Application of the boundary conditions
    that matches the value and radial derivative of the metric function
    at $r=r_{b}$ for $\Phi \neq 0,$ but where anisotropy compensates
    the charge.  The parameter values are
    $\rho_{c}=\un{1\times 10^{18}\,kg \cdot m^{-3}}, r_b = \un{1
      \times 10^{4}\,m}$
    and $\mu = 1$, and have been chosen to mimic values thought
    to be present in actual compact stars.}
\label{ns.fig:MetricAniCha}
 \end{figure}

In this section solutions to the EMS where the electric charge and
anisotropic pressure are related to each other through the relation
\(\Delta = 2(q/x)^{2},\) are discussed.  From the arguments in
Section~\eqref{ns.sec:phys}, \(q = kr^{3}.\) This particular choice
simplifies the differential equation for the \(Y\) metric function
allowing for a solution analogous to the Tolman VII solution for \(Y\)
to be written in the form
\begin{equation}
  \label{ns.eq:YAnisotropisedCharge}
  Y(\xi) = c_{1} \cos{\left( \Phi \xi \right)} + c_{2} \sin{\left( \Phi \xi \right)}, \quad \text{with } \Phi = \sqrt{\f{a}{4}}.
\end{equation}
However this solution is fundamentally different from the Tolman VII
solution which was a solution to the Einstein's system of equation and
not the Einstein--Maxwell system.  There are a number of reasons for
this:
\begin{enumerate}
\item The charge in this system is non-zero, unlike the Tolman VII
  solution, where \(Q = 0.\)  
\item The presence of the anisotropic pressure in the solution means that
  \(\ppen\) is \emph{not} the same as the radial pressure \(p_{r}.\)
  This is clear since \(\Delta \neq 0.\)
\item Also, this solution will have to be matched to the
  Reissner-Nordstr\"om metric outside the sphere, as opposed to the
  Schwarzschild solution for the Tolman~VII solution.
\end{enumerate}
If these conditions are implemented, a fully-fledged new solution to
the EMS can be obtained.  Applying the  boundary
conditions~\eqref{ns.eq:BoundaryApp1} and~\eqref{ns.eq:BoundaryApp2},
at the vacuum--matter interface leads to:
\begin{itemize}
\item The condition on the derivative  
  \[\left. \deriv{Y}{\xi}\right|_{\xi=\xi_{b}} = \Phi \left[
    c_{2}\cos{(\Phi \xi_{b})} -c_{1}\sin{(\Phi \xi_{b})} \right] =
  \alpha, \]
  which can be rearranged to yield an equation for \(c_{1}\)
  and \(c_{2}\)
  in terms of previously defined constants:
  \(c_{2}\cos(\Phi \xi) -c_{1}\sin(\Phi \xi) = \alpha / \Phi\) and
\item The second condition on the function \(Y\):
\[Y(r_{b}) = c_{1}\cos{(\Phi \xi_{b})} + c_{2} \sin{(\Phi \xi_{b})} = \gamma,\]
\end{itemize}
Together this pair of equations can be solved for \(c_1\)
and \(c_2\) to give
\begin{align*}
  c_{2} &= \gamma \sin{(\Phi\xi_{b})} + \f{\alpha}{\Phi} \cos{(\Phi\xi_{b})}\\
  c_{1} &= \gamma \cos{(\Phi\xi_{b})} - \f{\alpha}{\Phi} \sin{(\Phi\xi_{b})},
\end{align*}
Figure~\ref{ns.fig:MetricAniCha}
which plots the metric functions demonstrates the effects of the
parameter \(k\)
which measures the magnitude of the electrostatic charge.  For large
values of \(k\)
the behaviour of the metric functions deviates significantly from the
\(k=0\) cases studied in Section~\ref{ns.sec:Ani}

 The complete solution where the anisotropic pressures and the charge
 compensate for each other thus becomes
\begin{multline}
\label{ns.eq:YphiCCA}
    Y(r) = \\
    \left( \gamma \cos{(\Phi\xi_{b})} - \f{\alpha}{\Phi} \sin{(\Phi\xi_{b})} \right) 
    \cos{\left( \f{2\Phi}{\sqrt{a}} \coth^{-1} \left( \f{1+\sqrt{1-br^2+ar^4}}{r^2\sqrt{a}}\right)  \right)} +\\
    + \left(\gamma \sin{(\Phi\xi_{b})} + \f{\alpha}{\Phi} \cos{(\Phi\xi_{b})} \right) 
    \sin{\left(\f{2\Phi}{\sqrt{a}} \coth^{-1} \left( \f{1+\sqrt{1-br^2+ar^4}}{r^2\sqrt{a}}\right) \right)},
  \end{multline}
which then allows us to write the matter variables \(\ppen\) and \(p_{r}\) as
\begin{multline}
  \label{ns.eq:PhiPprCCA}
\kappa p_{r}(r) = \f{2\kappa\rho_{c}}{3} - \f{4}{5}\left( \f{\kappa \rho_{c}\mu}{r_{b}^{2}} - k^{2}\right) r^{2} -\kappa\rho_{c} \left[ 1 - \mu \left( \f{r}{r_{b}}\right)^{2}\right] + 4 \Phi\sqrt{1-br^{2}+ar^{4}} \times \\
\times \left\{ \f{ \left[ \gamma \sin{(\Phi\xi_{b})} + \f{\alpha}{\Phi} \cos{(\Phi\xi_{b})} \right] \cos{(\Phi \xi)} - \left[ \gamma \cos{(\Phi\xi_{b})} - \f{\alpha}{\Phi} \sin{(\Phi\xi_{b})} \right] \sin{(\Phi \xi)} }{\left[ \gamma \sin{(\Phi\xi_{b})} + \f{\alpha}{\Phi} \cos{(\Phi\xi_{b})}\right] \sin{(\Phi \xi)} + \left[\gamma \cos{(\Phi\xi_{b})} - \f{\alpha}{\Phi} \sin{(\Phi\xi_{b})} \right]\cos{(\Phi \xi)}} \right\},  
\end{multline}
and 
\begin{equation}
  \label{ns.eq:PhiPptCCA}
  \ppen(r) =p_{r} - \Delta = p_{r} - 2k^{2}r^{2}   
\end{equation}
The intermediate variables in the above expressions for this case can
be given in terms of the free parameters by :
\begin{align*}
  \alpha &= \f{\left( \kappa\rho_{c}(5 -3\mu) - 12k^{2}r_{b}^{2} \right)}{60}, &\quad& \Delta(r) = 2k^{2} r^{2} = \f{qr}{2k}, \\
  \gamma &= \sqrt{1+ \f{\kappa \rho_{c}r_{b}^{2}(3\mu-5)}{15} -\f{k^{2}r_{b}^{2}}{5}}, &\quad& \Phi^{2} = \f{1}{4}\left(\f{\kappa\rho_{c}\mu}{r_{b}^{2}} - k^{2} \right), 
\end{align*}
which completes the solution.  As can be seen the solution could be
expressed in terms of \(q,\)
or \(\Delta\)
exclusively, since these two functions are not independent in this
particular solution.  As with the previous examples, an explicit
equation of state could be obtained since it requires a
straightforward inversion of the density function's dependence on
\(r.\)
The total mass and charge of the object modelled by this solution is
obtained through~\eqref{ns.eq:Mass+Charge}, and for this particular
case, these equations simplify to
\begin{equation}
  \label{ns.eq:M+CCCA}
M = 4\pi\rho_{c}r_{b}^{3} \left( \f{1}{3} - \f{\mu}{5} \right) +\f{k^{2}r_{b}^{5}}{10}, \quad \text{and,} \quad  Q = kr_{b}^{3}.  
\end{equation}
The last equation can be used to determine the charge density
\(\sigma(r), \) since from~\eqref{ns.eq:Mass+Charge}, 
\begin{equation*}
  \int_{0}^{r_{b}} \bar{r}^{2} \d r \left[ 4\pi \sigma(\bar r) \sqrt{Z(\bar r)} \right] = Q = kr_{b}^{3} = \int_{0}^{r_{b}} \bar{r}^{2} \d r \left[ 3 k \right].
\end{equation*}
Direct comparison of terms yields the charge density
\begin{equation}
  \label{ns.eq:sigmaCCA}
  \sigma(r) = \f{3k}{4 \pi \sqrt{Z(r)}}.
\end{equation}
This completes the solution for this case.  Turning to the case where
both charge and anisotropic pressure exist independently of each
other, requires a thorough analysis of the different combinations of
charge and anisotropic pressure, and how these conspire to change the
character of the differential equation.
 
\subsection{The $\Phi^{2} = 0$ case}\label{ns.sssec:phi0}
If \(\Phi^{2} = 0\) in equation~\eqref{ns.eq:YdiffXi}, then
\(2k^{2} = a + \beta,\) and the solution for \(Y\) is simply
\(Y = c_{1} +c_{2} \xi,\) with \(c_{1}\) and \(c_{2}\) constants of
integration.  Applying boundary conditions on this solution then
results in
\[\left. \deriv{Y}{\xi} \right|_{\xi=\xi_{b}} = c_{2} = \alpha :=
\f{1}{4} \left( \f{\kappa \rho_{c}}{3} - \f{3\kappa\rho_{c}\mu}{11} -
  \f{4r_{b}^{2}\beta}{11} \right),\] and
\[ \left. Y \right|_{\xi = \xi_{b}} = c_{1} + c_{2}\xi_{b} = \gamma :=
\sqrt{1+r_{b}^{2}\kappa\rho_{b}\left( \f{2\mu}{11} - \f{1}{3}\right)
- \f{\beta r_{b}^{4}}{11}},\]
which can be solved together algebraically to give the value of
\(c_{1}.\) This completes the solution for \(Y\) in this particular case.

The \(Z\)
metric function is still fixed by the Tolman assumption:
\(Z = 1 -br^{2} + ar^{4}.\)
In this particular case \(a\) and \(b\)  are given by:
\[a = \f{2}{11}\left( \f{\kappa \mu \rho_{c}}{r_{b}^{2}} -
  \f{\beta}{2}\right), \quad \text{and} \quad b =
\f{\kappa\rho_{c}}{3}. \]

\begin{figure}[h!]
  \centering
  \includegraphics[width=\linewidth]{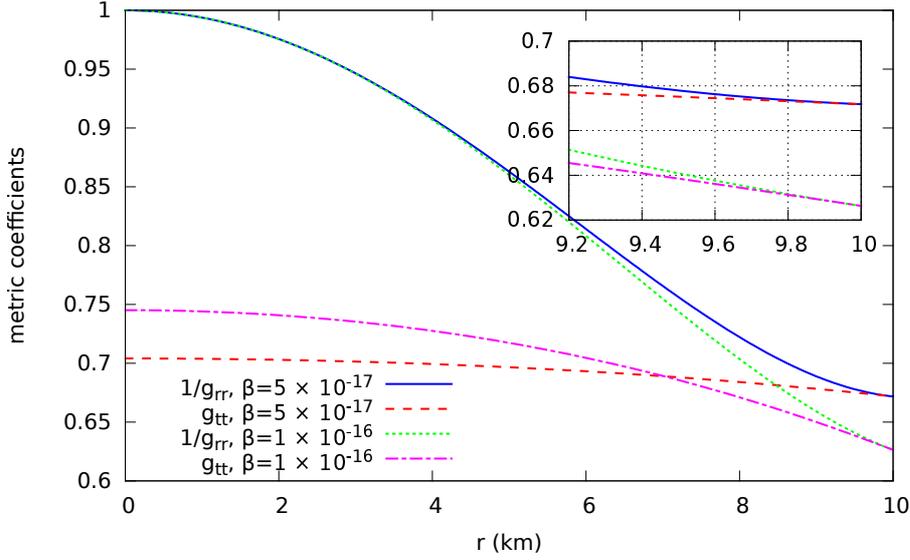}
  \caption[Matching the metric functions for $\Phi = 0$] {Application
    of the boundary conditions that match in the value and radial
    derivative of the metric function at $r=r_{b}$ for the $\Phi^{2} = 0$
    case. The parameter values are
    $\rho_{c}=\un{1\times 10^{18}\, kg \cdot m^{-3}}, r_b = \un{1
      \times 10^{4}\, m}$ and $\mu = 1.$ The parameter values are
    chosen to mimic the values these parameters would have had in compact real
    stars.}
\label{ns.fig:MetricAniChaPhiZ}
\end{figure}

Given the relation \(a+\beta = 2k,\)
only two parameters are arbitrary to completely specify the solution,
and here, \(\beta\)
is chosen in preference to \(k.\)
This feature, and the consistent matching of the boundaries is shown
in Figure~\ref{ns.fig:MetricAniChaPhiZ}.

Once the two metric functions are found,  all other quantities are
determined, in particular the radial pressure \(p_{r}\) is given by
\[p_{r} = \f{1}{\kappa} \left[ \f{4c_{2} \sqrt{1 -br^{2} + ar^{4}}
  }{c_{1}+c_{2}\xi} +2b -4ar^{2} \right] - \rho(r),\]
and the tangential pressure \(\ppen,\)
in turn is \(\ppen = p_{r} - \Delta/\kappa,\)
giving\[\ppen = p_{r} - \f{\beta x}{\kappa}.\]
The mass \(M\)
and charge \(Q\)
seen from the exterior, which are given by~\eqref{ns.eq:M+CCCA}
result in
\[ M = 4\pi \rho_{c}r_{b}^{3} \left( \f{1}{3} - \f{7\mu}{55}\right) +
\f{\beta r_{b}^{5}}{22}, \quad \text{and}, \quad Q = r_{b}^{3}
\sqrt{\left( \f{5\beta}{11} +
    \f{\kappa\mu\rho_{c}}{11r_{b}^{2}}\right)}.\]
This completes the solution for this particular specific case.

\subsection{The $\Phi^{2} < 0$ case}\label{ns.sssec:phiN}
Here \((a+\beta-2k^{2})/4 < 0,\)
and the solutions for \(Y\) can be written in the form
\[Y = c_{1}\cosh{\left(\Phi \xi \right)} + c_{2}\sinh{\left(\Phi \xi
    \right)}.\] In this particular case, one will not have a
simplification wherein the charge could be compensated completely by
the anisotropic pressure or mass, and one is then forced to deal with
all three contributions.  In this case \(2k^{2} > a+\beta,\) and this
leads to the belief that such a solution has very little chance of
being physical, since the charge contribution dominates the mass and
pressure.

Applying boundary conditions to this solution to obtain the values of
the constants \(c_{1}\)
and \(c_{2}\) through the same procedure as previously, leads to
\[\left. \deriv{Y}{\xi}\right|_{\xi=\xi_{b}} = \Phi \left[ c_{2}
  \cosh{(\Phi \xi_{b})} + c_{1} \sinh{(\Phi \xi_{b})}\right] =
\alpha, \]
and
\[Y(\xi_{b}) = c_{2}\sinh{(\Phi \xi_{b})} +c_{1}\cosh{(\Phi \xi_{b})}
= \gamma.\]

\begin{figure}[h!]
  \centering
  \includegraphics[width=\linewidth]{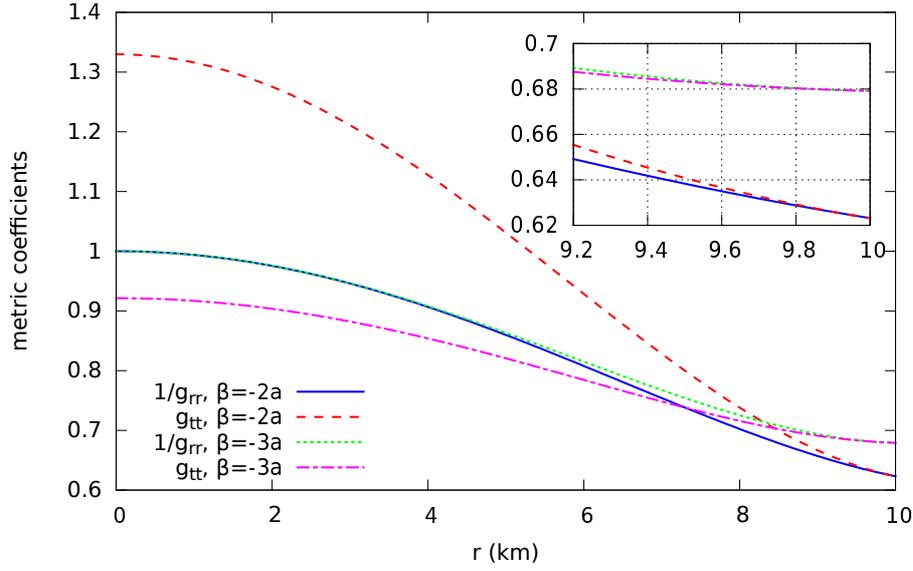}
  \caption[Matching the metric functions for $\Phi < 0$]{Application
    of the boundary conditions that match the value and radial
    derivatives of the metric function at $r=r_{b}$ for $\Phi^{2} < 0.$
    The parameter values are
    $\rho_{c}=\un{1\times 10^{18}\, kg \cdot m^{-3}}, r_b = \un{1
      \times 10^{4} \,m}$ and $\mu = 1.$ The parameter values are
    chosen to mimic values in actual compact objects}
\label{ns.fig:MetricAniChaPhiN}
\end{figure}

Then using a procedure very similar to that of previous sections, the
integration constants can be computed as
\begin{align}
  c_{2} &= \f{\alpha}{\Phi} \cosh{(\Phi\xi_{b})} - \gamma\sinh{(\Phi\xi_{b})},\\
  c_{1} &= \gamma \cosh{(\Phi\xi_{b})} - \f{\alpha}{\Phi} \sinh{(\Phi\xi_{b})}
\end{align}
We show the solutions up to the boundary \(r=r_{b}\) in
Figure~\ref{ns.fig:MetricAniChaPhiN}, and note that in this case we
need both \(\beta\) and \(k\) to completely specify one particular
solution.

\subsection{The $\Phi^{2} > 0$ case}\label{ns.sssec:phiP}
This implies \((a+\beta-2k^{2})/4 > 0,\) and now the solutions are
given
by\[Y = c_{1}\cos{\left(\Phi \xi \right)} + c_{2}\sin{\left(\Phi \xi
    \right)}.\] As in the previous case, the simplification wherein
the charge could be compensated completely by the anisotropic pressure
or mass, does not happen, and one is forced to deal with all three
parameters.  However the charge contribution will be less than the
mass and pressure anisotropy contribution (since
\(2k^{2} < a+\beta,\)) therefore this would seem to be the most
promising candidate for a new physically relevant solution.
Investigating this solution and the remaining ones, in detail will
allow conclusions about viability as a physical solution to be drawn.

\begin{figure}[h!]
  \centering
  \includegraphics[width=\linewidth]{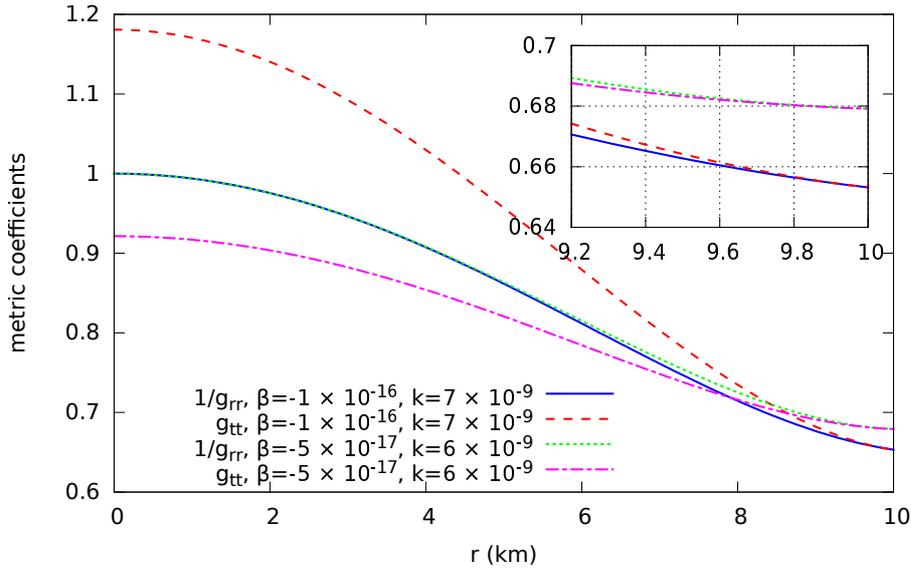}
  \caption[Matching the metric functions for $\Phi < 0$]{Application
    of the boundary conditions that match the value and radial
    derivative of the metric function at $r=r_{b}$ for $\Phi^{2} < 0.$ The
    parameter values are
    $\rho_{c}=\un{1\times 10^{18}\,kg \cdot m^{-3}}, r_b = \un{1
      \times 10^{4}\,m}$ and $\mu = 1$, and were chosen to mimic the
    values in compact objects}
\label{ns.fig:MetricAniChaPhiP}
 \end{figure}

Applying boundary conditions to this solution the values of the
constants \(c_{1}\) and \(c_{2}\) are obtained.
\begin{enumerate}
\item \(\left. \deriv{Y}{\xi}\right|_{\xi=\xi_{b}} = \Phi \left[ c_{2} \cos{(\Phi \xi_{b})} - c_{1} \sin{(\Phi \xi_{b})}\right] = \alpha, \) and 
\item \(Y(\xi_{b}) = c_{2}\sin{(\Phi \xi_{b})} +c_{1}\cos{(\Phi \xi_{b})} = \gamma.\)
\end{enumerate}
Then using a procedure similar to that of previous sections one
can obtain for the integration constants
\begin{align}
  c_{2} &=  \gamma\sin{(\Phi\xi_{b})} + \f{\alpha}{\Phi} \cos{(\Phi\xi_{b})},\\
  c_{1} &= \gamma \cos{(\Phi\xi_{b})} - \f{\alpha}{\Phi} \sin{(\Phi\xi_{b})}.
\end{align}
The solution inside the star for different \(\beta\) and \(k\) are
plotted in Figure~\ref{ns.fig:MetricAniChaPhiP}.

\section{Conclusions}
\label{ns.sec:New}
Beginning with the Tolman~VII ansatz that \(Z=1-br^{2} +a r^{4},\)
together with the addition of electric charge and anisotropic
pressures, it has been shown that a number of new solutions to the
Einstein--Maxwell system of equations can be constructed.  In addition
these solutions are all regular in the radial coordinate \(r.\) The
crucial step in constructing the new solutions is finding the function
\(Y(r),\) since the ODE leading to its solution is initially
non-linear and second order.

The choices stemming from requiring a physical solution determine the
radial dependence of \(Z\) and \(q.\) The functional form of \(Z\) is
chosen so that the integral of \(Z^{-1/2}(x)\) can be computed
analytically, and based on the Tolman~VII solution, the mass density
\(\rho\) generated by this assumption can be physically motivated.
The form for \(q,\) as was argued in Section~\ref{ns.sec:phys} was
chosen to be in line with the assumption for \(Z,\) since any other
choice would render the second order ODE~\cite{Iva02} for \(Y\) to be in the form
\begin{equation}
  \label{ns:eq.IvaODE}
(1-b x + ax^{2}) \sderiv{Y}{x} + \left( a x - \f{b}{2} \right) \deriv{Y}{x} + \left( \f{\Delta}{4x} + \f{a}{4} -\f{q^{2}}{2x^{3}} \right) Y = 0.
\end{equation}

Indeed one can only realistically modify the choice of the anisotropic
pressure in order to generate a new function.  The crucial step in
constructing solutions stems from equation~\eqref{ns.eq:YdiffXi},
which can then be converted by judicious choices to a situation where
the second derivative of \(Y\) with respect to \(\xi\) is proportional
to \(Y.\) The restriction of the radial dependence for the charge
\(q\) to be a cubic function leads to an ODE for \(Y\) in the form:
\[\sderiv{Y}{\xi} - \f{k^{2}}{2}Y + \f{Y}{4}\left( a +
    \f{\Delta}{x}\right) = 0,\] after the transformation to the
\(\xi\) radial variable as discussed in Section~\ref{ns.sec:ODE}.

Of course when pursuing uncharged solutions \(k\) (and \(q\)), vanish.
Then simple solutions can be obtained by an appropriate choice of
\(\Delta.\) However, if charged solutions are required, modifying the
terms in brackets in such a way as to keep a simple form for the \(Y\)
ODE, appears to be the best way to proceed.

The application of boundary conditions on the metric functions \(Y\)
and \(Z\) completed the closed form solutions in terms of parameters
that can be \emph{physically} interpreted.  Expressions for the matter
variables: the density \(\rho,\) the pressures \(p_{r}\) and
\(\ppen,\) the electric charge \(Q,\) and the mass \(M\) for the
models were also specified.  Issues such as the stability of the
solutions to radial perturbations and/or the regularity of the matter
variables have been investigated in~\cite{phdThesis} and will be
published subsequently.  The goal here is to provide a simple method
of generating some solutions to the EMS that begin with physically
motivated ans\"atze.  Following a similar analysis given
in~\cite{RagHob15} the expectation is that some of these solutions may
lead to exact analytic models of realistic compact stars.

This article should be regarded as the mathematical component of
solution finding, and model building for the solutions.  The physical
description of compact objects described by these solutions will be
forthcoming.

\begin{acknowledgements}
  We kindly thank Dr. Eugene Couch with whom we had discussions about
  the possibility of extending the main approach to new solutions.
  Funding for this research was obtained from a Natural Sciences and
  Engineering Research Council of Canada (NSERC) Discovery Grant.
\end{acknowledgements}
\nocite{Bay78} 
\bibliographystyle{spmpsci}
\bibliography{bibliography}

\end{document}